\documentstyle[11pt,fleqn,aaspptwo,tighten]{article}
\newcommand{\sfRO}{\renewcommand{\baselinestretch}{1.1}\begin{small}}
\newcommand{\tfRO}{\end{small}\renewcommand{\baselinestretch}{1.6}}
\newcommand{\Sec}{\mbox{{\rm Sec. }}}

\newcommand{\etal}{\mbox{{\it et al. }}}

\newcommand{\disp}[2]{\mbox{$\sigma#2_{{\rm #1}}$}}

\newcommand{\Kz}[1]{\mbox{$K_{\rm z}{#1}$}}
\newcommand{\qrho}{\mbox{$q_{\rho}$}}
\newcommand{\rhoN}{\mbox{$\rho_0$}}
\newcommand{\Rc}{\mbox{$R_{\rm c}$}}
\newcommand{\Rcsq}{\mbox{$R_{\rm c}^2$}}

\newcommand{\pmt}{\mbox{$\pm \;$}}

\newcommand{\rtp}[1]{\mbox{$^{#1}$}}

\newcommand{\Ht}{\mbox{${\rm H_2}$}}
\newcommand{\HI}{\mbox{{\rm H \footnotesize{I} }}}

\newcommand{\Msun}{\mbox{$M_{\odot}$}}

\newcommand{\hR}{\mbox{$h_{\rm R}$}}

\newcommand{\ze}{\mbox{$z_{\rm e}$}}

\newcommand{\LSpcsq}{\mbox{$L_{\odot}{\rm pc}^{-2}$}}
\newcommand{\MSpcsq}{\mbox{$M_{\odot}{\rm pc}^{-2}$}}

\newcommand{\MoverL}[1]{\mbox{$M/{\cal L}_{{\rm #1}} \;$}}
\newcommand{\kms}{\mbox{${\rm km \;s}^{-1}$}}

\newcommand{\add}{\mbox{$^{\rm o} \!\!.$}}

\newcommand{\ad} {\mbox{$^{\rm o} \;$}}
\newcommand{\VLAnote}{\footnote{The VLA of the 
National Radio Astronomy Observatory is a facility of the National
Science Foundation operated under cooperative agreement by Associated 
Universities, Inc.} \ }
\newcommand{\QrhoNote}{\footnote{A spheroidal system with 
shortest-to-longest axis ratio $c/a \; (q_{\rho})$ of the density contours
has a shape E$n$, with $n$ such that $q_{\rho}=1-n/10$}}
\newcommand{\SOTON}{Dept.  of Physics, University of Southampton, 
Southampton S17 1BJ, U.K.}
\chardef\Isp="10

\begin{document}

\title{The Highly Flattened Dark Matter Halo Of NGC 4244}

\author{Rob P. Olling}
\affil{Columbia University, New York, \\
now at \SOTON \\
olling@astro.soton.ac.uk}
 
\begin{center}
\vspace*{2cm}
To be published in the Aug. 1996 issue of the Astronomical Journal
\end{center}
 
\authoraddr{\SOTON}


\begin{abstract}

In a previous paper (Olling 1995, \aj, 110, 591) a method was developed
to determine the shapes of dark matter halos of spiral galaxies from an
accurate determination of the rotation curve, the flaring of the gas
layer and the velocity dispersion in the \HI.  Here this method is
applied to the almost edge-on Scd galaxy NGC 4244 for which the
necessary parameters are determined in the accompanying paper (AJ, Aug. 
1996).

The observed flaring of the \HI beyond the optical disk puts significant
constraints on the shape of the dark matter halo, which are almost
independent of the stellar mass-to-light ratio.  NGC 4244's dark matter
halo is found to be highly flattened with a shortest-to-longest axis
ratio of $0.2_{-0.1}^{+0.3}$. If the dark matter is disk-like, the data
presented in this paper imply that the vertical velocity dispersion of
the dark matter must be 10\% - 30\% larger than the measured tangential
dispersion in the \HI.  Alternatively, the measured flaring curve is
consistent with a round halo if the gaseous velocity dispersion
ellipsoid is anisotropic.  In that case the vertical dispersion of the
gas is 50 - 70\% of the measured tangential velocity dispersion.

\end{abstract}

\sfRO
\section{INTRODUCTION}
\label{sec-Introduction}

Although rotation curves of spiral galaxies have been used as evidence
for the presence of dark matter (DM), little is known about the nature,
extent and actual distribution of the DM halos (e.g., van Albada \etal
1985; Lake \& Feinswog 1989).  As measurements of the equatorial
rotation curve probe the potential in only one direction, they provide
no information about the shape of the DM halos. 

In a previous paper, Paper I (Olling 1995), a method was developed to
determine the shape of the dark matter halo from the gaseous velocity
dispersion and the radial variation of the thickness of gas layers
(flaring).  This is accomplished by comparing the measured flaring with
that expected from a self-gravitating gaseous disk in the axisymmetric
potential due to the stellar disk and (flattened) DM halo.  The method
of calculation which produces the most reliable model flaring curve is
the {\em global} approach, in which the potential is calculated from the
{\em total} mass distribution of the galaxy (Paper I).  In this paper
the method is applied to the galaxy NGC 4244 for which the basic
parameters were determined in the accompanying paper (Olling 1996,
hereafter Paper II).

In the past, several methods have been used to determine the shapes of
dark matter halos.  Hofner \& Sparke (1994) analyzed the warping
behavior of \HI disks and concluded that only one (NGC 2903) of the five
systems studied requires a DM halo as flattened as E4\QrhoNote.  On the
other hand, in studies of polar ring galaxies (Sackett \& Sparke 1990;
Sackett \etal 1994; Sackett \& Pogge 1995) substantially flattened DM
halos are found (E6-E7 for NGC 4650A, E5 for A0136-0801).  The shape of
the dark halo of the Milky Way has been estimated (E0 - E7) from the
kinematics of extreme Population II stars (Binney \& Ostriker 1987b;
Sommer-Larsen \& Zhen 1990; van der Marel 1991; Amendt \& Cuddeford
1994).  Steiman-Cameron \etal (1992) used the dynamics of the precessing
dusty disk of the S0 galaxy NGC 4753 to infer the shape of its DM halo
(E1). 

Van der Kruit (1981) pioneered the use of flaring measurements to
determine the mass of stellar disks, and found that the scale length of
the total matter (luminous plus dark) was similar to the scale length of
the light distribution and concluded that the mass-to-light ratio does
not vary significantly with radius.  Since mass models where all the
mass is concentrated in the disk require large radial changes in the
mass-to-light ratio (\MoverL{}) to reproduce the observed flat rotation
curves, the DM halo is not as highly flattened as the stellar disk.  A
similar conclusion was reached for NGC 4565, NGC 891 (Rupen 1991), and
the Milky Way (Knapp 1987; Merrifield 1992; Malhotra 1994, 1995).  As we
can measure the light distribution of the stellar disk only and have no
prior knowledge of the mass-to-light ratio of stellar disks, the
relative contributions of luminous and dark matter to a given rotation
curve are not known.  Models of galaxy formation which include simple
estimates of the star formation history (e.g., Larson \& Tinsley 1978)
predict \MoverL{}'s which are in the range of values derived from
observed rotation curves (e.g., Rubin \etal 1985; Athanassoula \etal
1987; Bottema 1988; Broeils 1992, Chap.  12), but different star
formation histories can easily result in mass-to-light ratios which
differ by a factor of a few (e.g., Tinsley 1981; Worthey 1994; de Jong
1995, Chap.  4).  An upper limit to the mass-to-light ratio, and hence a
lower limit to the amount of DM, can be obtained by assuming that the
peak of the observed rotation curve is due to the stellar disk only, the
so called maximum-disk hypothesis (van Albada \& Sancisi 1986).  Other
\MoverL{} estimators have been used (van der Kruit 1981; Efstathiou
\etal 1982; Athanassoula \etal 1987; Martinet 1988; Bottema 1993),
typically yielding a stellar disk 50\% - 100\% as massive as the
``maximum-disk''.  Employing the maximum-disk model, Broeils (1992,
1995) finds that the ratio of dark to luminous matter lies between 0.5
and 10, where the larger values are found in low mass systems. 
Unfortunately, the flaring data presented in this paper cannot be used
to constrain the mass of the stellar disk of NGC 4244 (\Sec
\ref{sec-The.Stellar.Mass.Distribution}) beyond the limits imposed by
the shape of the rotation curve. 

In a study similar to the present one, Rupen (1991) analyzed high
resolution \HI observations of two edge-on galaxies, NGC 4565 and NGC
891.  For these galaxies he finds that the flaring data imply stellar
disks with masses two-third and one-half as massive as a
``maximum-disk'', consistent with van der Kruit's (1981, 1988) results
for NGC 891.  In addition, Rupen found that non-thermal pressure
gradients ($ \frac{dP_{\rm nt}}{dz}$) need probably not be included in
the equation of hydrostatic equilibrium
(\ref{eq:eqn.of.hydrostatic.equilibrium}).  The fact that NGC 891 and
NGC 4565 have similar luminous mass distributions, similar gas layer
widths and similar thermal pressure gradients ($ \frac{dP_{\rm
th}}{dz}$) but are estimated to have very different non-thermal pressure
gradients ($\left.  \frac{dP_{\rm nt}}{dz} / \frac{dP_{\rm th}}{dz}
\right|_{\rm NGC 4565} \approx \frac{1}{2} \; ; \; \left.  \frac{dP_{\rm
nt}}{dz} / \frac{dP_{\rm th}}{dz} \right|_{\rm NGC 891} \approx 4 $)
argues against the importance of non-thermal pressure gradients in
balancing the vertical force.

Below I summarize the relevant properties of NGC 4244 as determined in
Paper II and references therein.  The current starformation rate in NGC
4244 is low.  Apart from a region between 6.5 and 8 kpc to the
south-west, the radial light profile is well represented by an
exponential distribution: extinction is probably not very important. 
The properties of the neural hydrogen layer were derived from sensitive
high resolution VLA\VLAnote observations.  The total \HI distribution is
fairly regular and shows a gentle warp.  Several strong \HI surface
density variations are seen in the inner disk.  Beyond the optical disk
the surface density on both sides of the galaxy is similar, validating
the use of the average surface density profile in the model
calculations.  The same holds for the gaseous velocity dispersion, and
the inferred inclination angle.  Superimposed upon a gradual increase of
the gas layer width, the derived thickness of the gas layer shows a
characteristic increase/decrease on the inside/outside of spiral arms. 
These features are probably not real but are likely an artifact (induced
by spiral arm streaming motions) of the thickness-and-inclination
derivation technique (Paper II).  NGC 4244 is an intermediate mass
system with a rotation speed of about 100 \kms.  The large inclination
($\sim$ 84\ad5) of NGC 4244 in combination with the symmetric surface
density distribution, allowed for the accurate determination of the
rotation curve.  NGC 4244 is one of the few galaxies to show, on both
sides, the characteristic drop in rotation velocity beyond the optical
disk.  This decline in rotation speed is more typical of luminous fast
rotators (Casertano \& van Gorkom 1991).  The slow rise in the inner
part and the decline of the rotation curve beyond the optical disk limit
the mass-to-light ratio of NGC 4244 to 50\% - 100\% of the maximum-disk
value.  Due to the limited extend of the measured rotation curve the
dark halo's core radius (\Rc) and central density (\rhoN) could be
determined to within a factor of 50 and 900 only, respectively.  This
large uncertainty in the dark-halo's density distribution does not
preclude the determination of the halo's shape (Paper I). 

In \Sec \ref{sec-The.Mass.Model} I describe how, in conjunction with
self-consistent mass models, the observed flaring can be used to
constrain the shape of the dark matter halo.  The shape of the halo is
then determined in \Sec \ref{sec-The.Shape.of.the.Dark.Halo}.  Some
caveats related to various step along the way are discussed in \Sec
\ref{sec-Caveats}.  The implications of the highly flattened halo for
the amount and nature of dark matter are discussed in \Sec
\ref{sec-Discussion}.

%
%
\section{THE MASS MODEL}
\label{sec-The.Mass.Model}

The dependence of the thickness of the gas layer upon the shape of the
dark matter halo is described in Paper I.  Here I will only present four
of the most salient features.  That the thickness of a gas layer depends
on the axial ratio ($\qrho=c/a$) of a dark halo can be easily
understood.  For an axisymmetric ellipsoidal dark matter distribution
with fixed \Rc \ and \rhoN \ [like Eq.  (\ref{eq:eqn.for.rho.DM})], the
mass enclosed by an ellipsoid of axial ratio \qrho, and hence the
rotation speed, decreases with decreasing \qrho.  Thus, for a given
observed rotation curve, the densities of flattened halos have to be
larger than for a round halo .  The vertical force, proportional to the
mass enclosed (Gauss' theorem), increases as well so that flattened
halos have thinner gas layers.  The second important result obtained in
Paper I is that the mass-to-light ratio (or equivalently, the mass) of
the stellar disk has virtually no influence on the thickness of the gas
layer beyond the stellar disk.  Given the fact that the dark matter
density distribution is poorly known (van Albada \etal 1985; Lake \&
Feinswog 1989; Paper I; Paper II) this may seem surprising at first. 
However, close to the plane the vertical force is roughly proportional
to the radial force\footnote{$\Kz = z/\sqrt{z^2 +R^2} \; F_{\rm tot}
\approx z/R \; F_{\rm R} \propto \frac{z}{R} \; V^2_{\rm obs}$} which
is, by construction, the same for all disk-halo combinations
that reproduce the observed rotation curve: beyond the optical disk the
dependence on halo flattening dominates. Thirdly, the self-gravity
of the gas can contribute significantly to the vertical force beyond 2 -
3 optical scale lengths.  And finally, the model gas layer widths are
best calculated using the global approach, in which the vertical force
is determined from the global mass distribution of the galaxy.

Comparing the thickness measurements beyond the optical disk with model
flaring curves, calculated for a series of mass models with varying halo
flattening, yields the flattening of the DM halo.  The basic assumptions
are that the vertical force $\Kz(z)$ is balanced by gradients in the
thermal pressure ($P_{\rm th}$) and that the gaseous velocity dispersion
ellipsoid is round and independent of height above the plane, so that
the vertical Jeans equation [e.g., Paper I, Eq.  (1)] can be
approximated by the equation of hydrostatic equilibrium:

\begin{eqnarray}
\frac{dP_{\rm th}}{dz} &=& \rho_{\rm gas} \Kz  \; \; .
\label{eq:eqn.of.hydrostatic.equilibrium}
\end{eqnarray}

\noindent Input to these mass models are the gaseous velocity
dispersion, the observed rotation curve, and the density distributions
of the dark and luminous matter.  Below I describe how to arrive at
these density distributions.

Optical studies of edge-on spiral galaxies indicate that the vertical
scale height of the stellar disk is constant as a function of galactocentric
distance (van der Kruit \& Searle 1981a,b 1982a,b; Shaw \& Gilmore 1990;
Barteldrees \& Dettmar 1994), although there are theoretical and
observational indications that the shape of the vertical distribution
may change with radius (Burkert \etal 1992 and van Dokkum \etal 1994,
respectively).  Here a constant vertical scale height is assumed.  For
the vertical distribution I use an exponential as well as a
sech$^2(z/z_0)$ distribution.  The effects of a different choice for the
vertical stellar distribution are discussed in \Sec
\ref{sec-The.vertical.distribution}.  To avoid extinction features
(Paper II), a fit to the stellar radial surface brightness distribution
is used: $L(R) = L(0) \exp{(-R/\hR)}$ for $R \leq$ 10 kpc, followed by a
linear decline till 11 kpc and $L(R)=0$ beyond 11 kpc.  A constant
mass-to-light ratio (\MoverL{B}) is assumed.  The stellar density
distribution is thus given by the product of the mass-to-light ratio and
the adopted radial and vertical distributions. 

The gaseous surface density used is the sum of the \HI surface density
distribution, a crude estimate of the \Ht \ surface density
distribution, and 25\% He by mass (Paper II).  The {\em volume} density
is found from $\Sigma_{\rm gas}(R)$, \disp{gas}{(R)} and the total
potential, via the equation of hydrostatic equilibrium.  Details of this
procedure can be found in Paper I.  A Gaussian fit to this density
distribution yields the model gas layer widths (but note that deviations
from the Gaussian distribution occur inside the optical disk, see
footnote 4 in Paper I).  In \Sec \ref{sec-The.Shape.of.the.Dark.Halo}
these model widths will be compared to the measured flaring curve.

The dark halo volume density is parameterized as a flattened isothermal
sphere with core radius $R_{\rm c}(\qrho)$, central density
$\rho_0(\qrho)$, and flattening $\qrho(=c/a)$ (e.g., Sackett \& Sparke
1990; Sackett \etal 1994; Paper I):

\begin{eqnarray}
\hspace*{-3mm}
\rho(R,z;\qrho) &=& 
   \frac{ \rho_0(\qrho) \; \Rcsq(\qrho)}{\Rcsq(\qrho) + R^2 + (z/\qrho)^2} \; .
   \label{eq:eqn.for.rho.DM}
\end{eqnarray}

\noindent The dependency of \Rc \ and $\rho_0$ upon \qrho \ is such that
the rotation curve of any flattened density distribution described by
Eq.  (\ref{eq:eqn.for.rho.DM}) is practically indistinguishable from
its round equivalent (Paper I).  In the flat rotation curve regime, the
thickness of the gas layer in a density distribution like Eq. 
(\ref{eq:eqn.for.rho.DM}) is given by:

\begin{eqnarray}
\hspace*{-3mm}
FWHM_{\rm z}(R) &\approx& 2.35 \;
    \left( \frac{\disp{gas}{}}{V_{\rm rot,max}} \right) \; \times
    \nonumber \\
\hspace*{-3mm} && \hspace*{-5mm}
    \sqrt{\frac{2.4\;q}{1.4+q}} \; \; \sqrt{R^2 + \Rc^2} \; \; ,
    \label{eq:FWHM.z.gas.DM}
\end{eqnarray}

\noindent with $V_{\rm rot,max}$ the asymptotic rotation speed of the DM
halo [Paper I, Eq. (D31)]. 

\begin{figure}
\plotfiddle{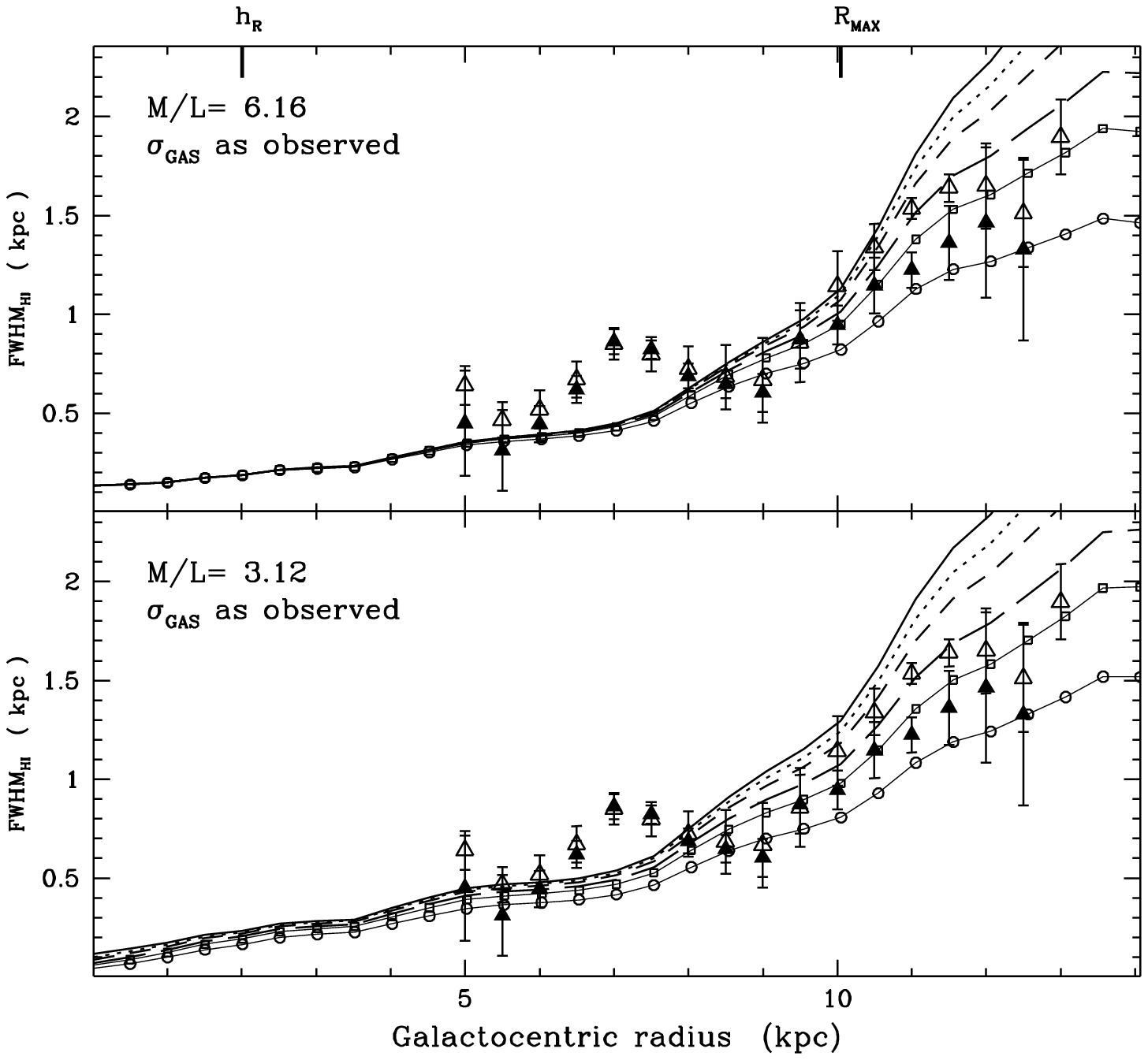}{10.0cm}{0}{100}{90}{-305}{-225}
\caption[Comparing the observed and the model flaring curves]{
\label{fig:derive.the.flattening}
The measured widths (open and filled triangles with error bars, for the
fixed and varying inclination case described in Paper II, respectively)
together with the model widths corresponding to different halo
flattenings $q_{\rho}$ are plotted: \qrho=1.0, 0.7, 0.5, 0.3, 0.2, and
0.1 from top to bottom.  The maximum-disk (top panel) and the
minimum-disk (bottom panel) disk-halo decompositions are presented (the
structural parameters are given in the caption of Fig.  15 of Paper II). 
Both the truncation of the stellar disk (at R$_{\rm MAX}$) and the
gaseous self-gravity have been included in the model calculation.  The
measured gaseous velocity dispersions (Paper II) were used in the model
calculations.  The observed flaring curve is the average of the
north-eastern and south-western sides.  Deemed unreliable in the inner 5
kpc (Paper II), the measurements from this region are not plotted.  All
plotted points are essentially independent measurements.  Beyond the
optical disk, the thickness of the gas layer is strongly influenced by
the shape of the dark halo.  Using both inclination cases I find: \qrho
= 0.2$_{-0.1}^{+0.3}$.  The mass-to-light ratio of the stellar disk is
not constrained by these flaring measurements. 
}
\end{figure}

As the mass models described above are fully specified by two
independent free parameters, the stellar mass-to-light ratio and the
flattening of the DM halo, the flaring behavior can be investigated as a
function of these parameters.  For every choice of the stellar
\MoverL{}, the DM core radius and central density are uniquely
determined by two points on the observed rotation curve (e.g., at 2.3\hR
\ and 8\hR, see Paper I, Appendix B).  However, the predicted flaring
beyond the optical disk is essentially independent of the choice of
\MoverL{} (Paper I; Fig.  \ref{fig:derive.the.flattening}).  The shape
of the DM halo is not a parameter in such rotation curve decompositions,
although different halo flattenings require somewhat different values of
\Rc \ and \qrho \ [Paper I, Eqs.  (B8) and (B9)].  In Paper II it was
found that the DM-halo parameters are not well constrained: possible
core radii range from 0.3 to 12 kpc and central densities between 1.3
and 0.001 \Msun/pc$^3$ are allowed; small \MoverL{}'s correspond to
small core radii and large halo densities.  Again, notwithstanding this
enormous allowed range in halo parameters, the thickness of the gas
layer in such disk-halo models (with the same flattening) is very
similar. 

The Dark Matter halo model [Eqn.  (\ref{eq:eqn.for.rho.DM})] is not
unique (e.g., van Albada \etal 1985; Lake \& Feinswog 1989; Navarro,
Frenk \& White 1996).  However, all DM density distributions with
similar rotation curves share the feature that flatter distributions
have larger vertical forces and hence thinner gas layers.  Thus, the
flattening determined in this paper is an indicator of the true
flattening of the true DM density distribution.  In the discussion (\Sec
\ref{sec-Discussion}) I investigate the extreme possibility that the
dark matter is disk-like.

\subsection{The Shape of the Dark Halo}
\label{sec-The.Shape.of.the.Dark.Halo}

Now we can compare the calculated gas scale heights for several mass
models with various \MoverL{} and \qrho \ with the observed widths (Fig. 
\ref{fig:derive.the.flattening}).  The truncation of the stellar disk,
the gaseous self-gravity and the measured gaseous velocity dispersions
have been included in the model calculations.  Inspecting this figure we
notice general correspondence between the observations and the model
flaring curves, except for the 7 kpc hump, where the thickness of the
gas layer increases by a factor 2.7 within 1 kpc (on both sides of the
galaxy).  Similar features are seen in the Milky Way (Knapp 1987) and
NGC 891 (Rupen 1991) but are absent in NGC 4565 (Rupen 1991).  The
steep thickness gradients in NGC 891 and NGC 4244 seem to occur just
inside and just outside \HI surface density enhancements.  Such sudden
changes in apparent thickness can arise from streaming motions induced
by spiral structure (Paper II, Appendix C.).

For the best-fit halo flattening, the maximum-disk and the minimum-disk
model curves seem to represent the flaring data equally well (upper and
lower panel, respectively).  This is corroborated by the fact that the
model with the largest reduced $\chi^2$ lies within the 1-$\sigma$ bound
(i.e., $\chi^2 \leq \chi^2_{\rm min} + 1)$ of the model with the
smallest reduced $\chi^2_{\rm min}$.  The north-eastern, south-western,
and the average flaring curves all yield the same estimate for the
mass-to-light ratio.  Thus, for NGC 4244, the measured flaring does not
constrain the mass-to-light ratio of the stellar disk beyond the limits
set by the shape of the rotation curve. 

On the other hand, the flattening of the halo is much better
constrained.  For all stellar disk masses, the minimum $\chi^2$ is found
at the same halo flattening, with the same formal uncertainty.  The
uncertainty in the inclination determination in the outer parts of the
galaxy (Paper II) has only a small effect on the inferred DM-halo shape:
for the varying-inclination case I find \qrho=0.2 \pmt 0.1, the
fixed-inclination case yields \qrho=0.3$_{-0.1}^{+0.2}$.  There are
several reasons why I believe that the small decrease in inclination
beyond the optical disk is probably real: 1) it is seen on both sides of
the galaxy, 2) it is seen (with lower S/N) when using the low velocity
channels (8-14) only, as well as in the ``edge'' channels (15-20) only. 
Nevertheless, I include the fixed-inclination case in the final
solution: $\qrho = 0.2^{+0.3}_{-0.1}$.

%
%
\section{CAVEATS}
\label{sec-Caveats}

We have seen that the dark halo of NGC 4244 is significantly flattened. 
Below I discuss the main uncertainties in this result.

\subsection{The Gaseous Velocity Dispersion}
\label{sec-The.Gaseous.Velocity.Dispersion}

\begin{figure}
\plotfiddle{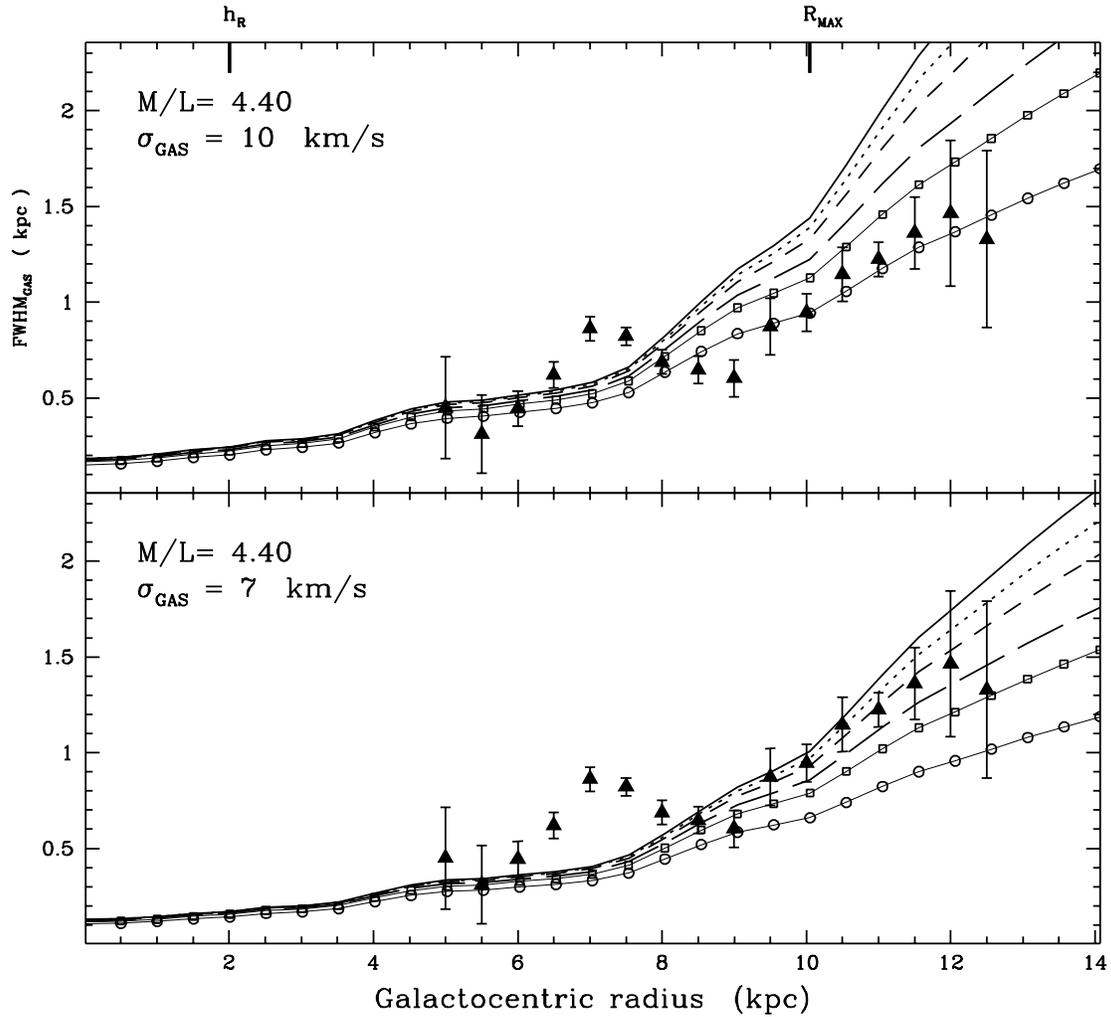}{11.5cm}{0}{100}{100}{-315}{-250}
\caption[The Halo flattening: its velocity dispersion dependence]{
\label{fig:Qrho.VelDis.dep}
The inferred halo flattening depends strongly upon the value of the velocity
dispersion.  The curves in the top panel (the same labeling of the
curves applies as in Fig.  \protect\ref{fig:derive.the.flattening}) are
calculated for the ``upper limit'' of the velocity dispersion as found
in the literature.  Similarly, the lower panel is calculated for the
``lower limit''.
}
\end{figure}

\begin{figure}
\plotfiddle{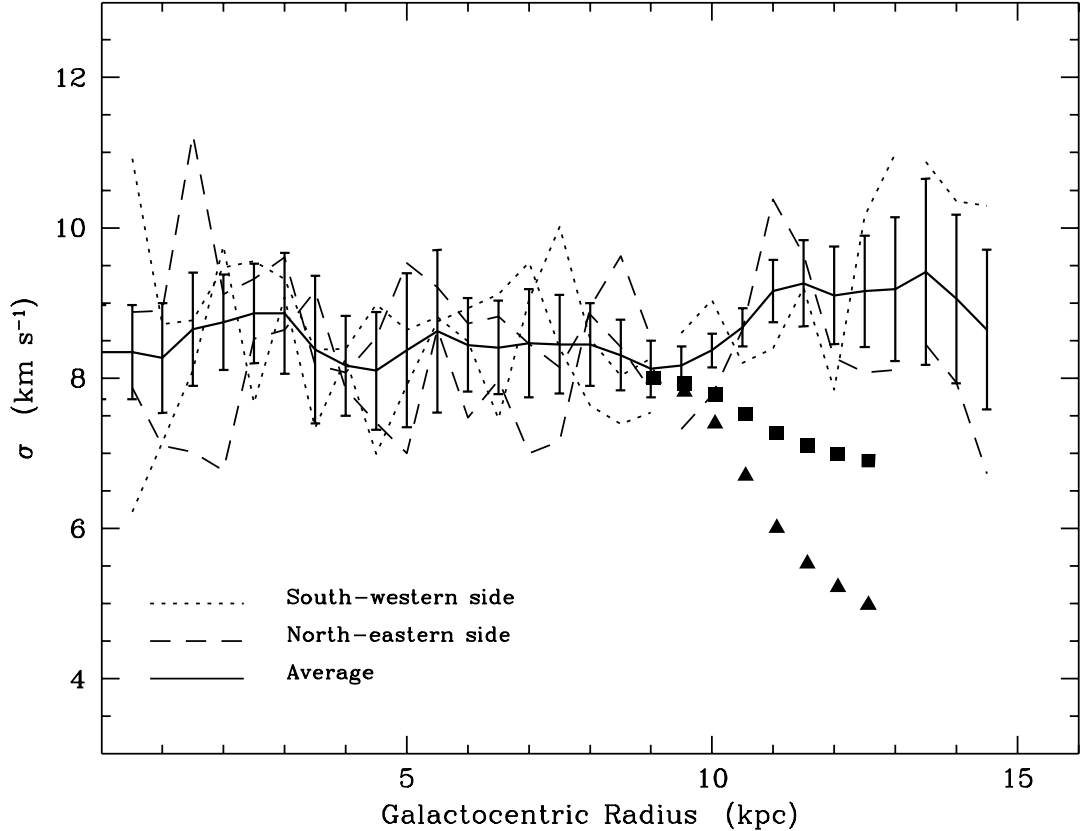}{11.5cm}{0}{100}{100}{-315}{-250}
\caption[Velocity dispersion required for a round DM halo]{
\label{fig:sigGAS.for.round.halo}
The radial variation of the average of the velocity dispersion (full
line with the error bars) as determined in Paper II is presented.  The
dotted lines are determined from the south-western side, the dashed
lines from the north-eastern side.  For all galactocentric radii, the
full line is the average of all curves shown.  The filled triangles are
the velocity dispersions required to make the model widths for a round
halo model consistent with the flaring measurements (see Fig. 
\protect\ref{fig:derive.the.flattening}: i.e., $\sigma_{zz}$ would
decrease by almost a factor two over 3 kpc).  To indicate the importance
of simultaneously determining the thickness and inclination, I included
the required $\sigma_{zz}$ in case an inclination had to be assumed in
determining the thickness of the gas layer (filled squares). 
}
\end{figure}

As the calculated thickness of the gas layer and the inferred halo
flattening depend, respectively, linearly and inversely quadratically on
the velocity dispersion of the gas\footnote{From Eq. 
(\ref{eq:FWHM.z.gas.DM}) it follows that $FWHM_{\rm z}^2 \propto
\sigma^2_{\rm gas} \; \qrho$.}, the velocity dispersion of the gas is an
important parameter in the analysis presented above.  This strong
dependence in illustrated in Fig.  \ref{fig:Qrho.VelDis.dep}, where I
compare model gas layer widths calculated for the lower and upper bounds
to the gaseous velocity dispersion found in other, more face-on, systems
(van der Kruit \& Shostak 1982; Dickey \etal 1990; see Kamphuis 1993,
Chap.  12, for a review).  It is striking that the shape of the halo
cannot be determined if no velocity dispersion {\em measurements} are
available: measuring the gaseous velocity dispersion is instrumental in
the determination of the shape of the dark matter halo.  From Fig. 
\ref{fig:sigGAS.for.round.halo} we can see that forcing the flaring
measurements to be consistent with a round DM-halo model requires a
radial decline of the velocity dispersion which is inconsistent with the
dispersion measurements.  Given that line-of-sight smearing effects are
unimportant and that the data from which the tangential velocity
dispersion was determined has a high signal to noise ratio ($S/N \ge 10$
for $R \le 14$ kpc; Paper II), it is highly unlikely that the measured
velocity dispersion curve is consistent with the velocity dispersion
needed to infer a round halo. 

The flaring data can be consistent with a round halo if the vertical
dispersion ($\sigma_{\rm zz}$) does not equal the measured tangential
dispersion ($\sigma_{\theta \theta})$.  The measured tangential
dispersion curve can thus be consistent with a round DM halo if
($\sigma_{\rm zz} / \sigma_{\theta\theta}$) decreases radially (down to
0.55 at the last measured point).  However, it is unclear whether such a
large anisotropy in the velocity dispersion of the \HI beyond the
optical disk is consistent with the observations.  For the molecular
component in the solar neighborhood this ratio is not well established:
$\sigma_{\rm zz} = 5.7 \pm 1.2$ for high latitude CO clouds (Blitz \etal
1984) while determinations of $\sigma_{\theta\theta}$ from tangent point
measurements vary between 3 \pmt 0.7 (Clemens 1985) and 7.8 \pmt 4
(Malhotra 1994). 

Assuming that an anisotropy in the gaseous velocity dispersion will be
reflected in the motions of young stars formed from this gas, proper
motion studies of OB associations may be informative.  On very small
scales ($\sim$ 0.5 pc) there is some evidence that the velocity
dispersion ellipsoid in the Orion Nebula Cluster is anisotropic
($\sigma_b/\sigma_l$ = 2.2/2.5 = 0.9 \pmt 0.1), while it is uncertain
how much this value is affected by the expansion of the cluster (Jones
\& Walker 1988).  On larger scales ($\sim$ 20 pc), the anisotropy of a
proper motion selected sample of stars is similar (2.9/3.2=0.9 \pmt 0.3;
McNamara \etal 1989), while a large proper motion and distance selected
sample (Warren \& Hesser 1977) covering a large volume (175$^3$ pc$^3$)
centered on Orion OB1 shows an increase of dispersion values but no
evidence for a velocity dispersion anisotropy (17.3/16.7 = 1.04 \pmt
0.2).  Mihalas \& Binney (1981, p.  423) report a small anisotropy in
the dispersion for O-B5 stars ($\sigma_{\theta\theta}/\sigma_{\rm zz} =
11/9 = 1.2$) with an unknown error.  In summary, the velocity dispersion
ellipsoid of young OB stars, and presumably of the interstellar medium,
is not significantly anisotropic in the solar neighborhood. 

Possibly the best way to determine the velocity dispersion ellipsoid, is
by comparing the velocity dispersions in a sample of face-on and edge-on
galaxies.  We (van Gorkom, Rupen \& Olling) are in the process of doing
so.

\subsection{The Determination of the Inclination}
\label{sec-The.Determination.of.the.Inclination}

Because the inferred halo flattening is proportional to the square of
the true width of the gas layer [Eq.  (\ref{eq:FWHM.z.gas.DM})], the
thickness of the gas layer has to be measured reliably (Paper II).  The
difference in the inferred \qrho \ between the thickness for the
fixed-inclination case (\qrho = 0.3$^{+0.2}_{-0.1}$) and the
free-inclination case (\qrho = 0.2 \pmt 0.1) exceeds the \qrho-error
resulting from the errors in the {\em measured} gaseous velocity
dispersions by a factor of two.

\subsection{The Stellar Mass Distribution}
\label{sec-The.Stellar.Mass.Distribution}
    
\subsubsection{The vertical distribution}
\label{sec-The.vertical.distribution}

Since the shape of the DM halo can be determined irrespective of the
actual mass of the stellar disk, its vertical distribution is
unimportant as well.  

\subsubsection{Does light trace mass ?}
\label{sec-Does.light.trace.mass}

In \Sec \ref{sec-The.Mass.Model} it was assumed that the stellar
mass-to-light ratio is constant throughout the galaxy.  As argued above,
a vertically varying mass-to-light ratio does not affect the inferred
halo flattening.  

A radial gradient in the stellar \MoverL{} would change the radial
distribution of stellar mass.  Consider the simple case where the true
radial scale length differs from the value used for \hR; this difference
would expand the horizontal scale of the {\em predicted} flaring curves
by a factor $h_{\rm R,true}/h_{\rm R,false}$, while the measured flaring
remains unchanged.  Thus, the usage of too long a radial scale length
leads to an overestimate of the halo flattening (too round a halo).  In
many early type spirals the scale length as determined from
near-infrared photometry is significantly shorter than the scale length
at visual wavelengths (Terndrup \etal 1994; Peletier \etal 1994). 
Extinction can not be the sole reason for this radial blueing since such
color gradients are found in face-on systems as well (de Jong \& van der
Kruit 1994; de Jong 1995).  In fact, it may be that a radially varying
star formation history (mean stellar age) in combination with a radial
metallicity gradient cause the observed color gradients (Josey \&
Arimoto 1992; de Jong 1995, Chapter 4).  However, such color gradients
are small\footnote{There are twelve systems in de Jong's (1995) sample
which are comparable to NGC 4244: in scale length, total luminosity and
total \HI mass.  For those systems I find $h_{\rm K-band}/h_{\rm
B-band}=1.02 \pm 0.06$, corresponding to small color gradients.} for
late type, low luminosity systems like NGC 4244.  Since extinction is
not very important in NGC 4244 and the current star formation is low,
the measured blue light probably traces mass, so that the inferred
DM-halo flattening is probably unaffected.

\subsection{Distance Dependence}

By combining the equations (D30) and (B16) from Paper I, it follows
that the model thickness, inside the stellar disk, is proportional to
$\sqrt{\hR \ze}$, equivalent to a linear distance dependence.  Likewise,
the the thickness of the gas layer beyond the optical disk is linearly
proportional to galactocentric radius [Eq.  (\ref{eq:FWHM.z.gas.DM})],
thus depending linearly on distance.  As the linear scale of the
thickness measurements depend linearly on distance as well, the flaring
gas layer method yields distance-independent values for the
mass-to-light ratio of the stellar disk and the shape of the DM halo.

\subsection{The Influence of the Extragalactic Radiation Field}
\label{sec-The.Influence.of.the.Extragalactic.Radiation.Field}

The \HI surface density profile of M33 (Corbelli \etal 1989) and the
major axis profile of NGC 3198 (van Gorkom 1993) show sudden change in
slope at column densities below a few times 10$^{19}$ cm\rtp{-2}. 
Maloney (1993) models the NGC 3198 case extensively and argues that
these slope changes are evidence for the extragalactic radiation field
(EgRF) ionizing the neutral hydrogen layer.  As as result, the degree of
ionization will increase with height above the plane as well, so that
width of the {\em neutral} gas layer (${\rm FWHM}_{\HI}$) will be
smaller than the thickness of the column of total hydrogen $({\rm
FWHM_{tot}})$, at low column densities.  Since the low column density
regime generally coincides with the region where the thickness of the
\HI layer is sensitive to the shape of the DM halo, I investigate the
effects of the EgRF on the width of the gas layer.  Inspection of
Maloney's Figs.  4b and 7, and comparing the densities of the NGC 3198
model with the measured densities in NGC 4244, indicates that the
thickness of NGC 4244's \HI layer may be underestimated by $\sim 15\%$
at 12 kpc.  However, Maloney's models were calculated for an EgRF,
$\Phi_{i,4}$, of $10^4$ photons cm\rtp{-2} s\rtp{-1} while the
2-$\sigma$ upper limit is three times larger (Vogel \etal 1995). 
Assuming that the effects of the radiation field are similar if the
ratio of critical density\footnote{In an analytic approximation of the
effects of the EgRF, Maloney (1993) finds that at total column densities
below the critical density, the whole hydrogen column is ionized, while
his detailed analysis indicates a rapid change of fractional ionization
for column densities near $N_c^{HI}$ [$=4.45 \; 10^{19}$
$\sqrt{\Phi_{i,4} \; {\rm FWHM_{HI}}}$ cm\rtp{-2}, adapted from
Maloney's Eqn (19)].} ($N_c^{HI}$) to \HI surface density is similar, I
used Maloney's NGC 3198 model to determine the dependence of the ratio
of observed to total thickness ($r_{\rm FWHM}$) on the ratio of critical
to \HI surface density ($r_{\Sigma}$)\footnote{With
$r_{\Sigma} = N_c^{HI}/\Sigma_{HI}, {\rm and} \; r_{\rm FWHM} = ({\rm
FWHM_{HI}/FWHM_{tot}})$, I find that the following relation holds:
$r_{\rm FWHM} \approx 1 + 2.57 \left( 1-
\exp{(-(r_{\Sigma}/1.35)^{2.2})} \right) \; \pm 0.015$ for $r_{\Sigma}
\le 3$, while for larger $r_{\Sigma}$-values $r_{\rm FWHM}$ decreases
again, similar to Maloney's Fig.  7.}.  With this relation it is
possible to correct the observed thickness for the effects of the
extragalactic radiation field, for any observed thickness and \HI column
density, and assumed EgRF strength.  For the radii of interest I find
that the thickness of the total hydrogen is $(3.4, 7.4, 18.4, 23.7,
35.4)\times \Phi_{i,4} \; \%$ larger than the measured thickness of the
\HI layer at 10.5, 11, 11.5, 12, and 12.5 kpc, respectively.  For small
values of the radiation field, $\Phi_{i,4} < 2$, the thickness
corrections are small so that inferred halo flattening is not affected. 
For stronger radiation fields ($\Phi_{i,4}$= 2-3) we find: \qrho = 0.5
\pmt 0.2.  In summary, the effects of the extragalactic radiation field
on the determination of \qrho \ are at most of the same order of
magnitude as the uncertainties in the gaseous velocity dispersion and
inclination.

\subsection{Non-Thermal Pressures Gradients}

The analysis presented in this paper assumes the vertical gravitational
force is balanced by thermal pressure gradients only.  It is estimated
that in the solar neighborhood the pressures due to cosmic rays,
magnetic fields and kinetic gas pressure are of comparable magnitude
(Spitzer 1978, p.  234).  The magnetic field of the Milky Way has a very
long radial scale length (Kulkarni \& Heiles 1988, p.  146) and could
provide significant pressure, even beyond the optical disk.  However,
such a stratified magnetic field is unstable (Parker 1966) and may
therefore not contribute to the vertical pressure balance.  Non-thermal
radio emission indicates that cosmic rays are closely related to sites
of star formation (Bicay \& Helou 1990), and do not extend much beyond
such regions.  With NGC 4244's radio continuum flux very low (Paper II)
it is expected that cosmic ray pressure
and/or magnetic field strength contributes little to the total pressure,
especially beyond the optical disk.  In the Introduction I summarized
observational evidence against the importance of non-thermal pressures
(Rupen 1991).  At any rate, if non-thermal pressures {\em are}
important, then an even larger DM-halo midplane density is required to
produce the observed flaring: additional non-thermal pressures require
an even flatter DM halo (see also Pfenniger \etal 1994, and references
therein).

\subsection{Combined systematic effects}
\label{sec-Combined.systematic.effects}

Having established that the velocity dispersion ellipsoid of the \HI is
probably round \\ (\Sec \ref{sec-The.Gaseous.Velocity.Dispersion}), I like
to stress once more the importance of determining the thickness of the
gas layer and the velocity dispersion accurately.  Furthermore, due to
projection effects, the derived thickness of the gas layer is quite
sensitive to the exact value of the inclination (for $i85\ad \; \pm \;
2\ad$; Paper II).  I illustrate this by comparing the inferred DM halo
flattenings for the case that gas layer widths are determined assuming a
constant inclination of 84\add5, and the case where the whole spectral
line cube is used to determine thickness and (varying) inclination
simultaneously.  If the galaxy were exactly edge-on or at an inclination
$\la$ 80\ad but larger than 60\ad, the uncertainty in the derived gas
layer width would be much less severe (Paper II).

For example, in the varying (fixed) inclination case the dispersion has
to be decreased by 6.2 (3.2) times the errors on the dispersion
measurements\footnote{Alternatively, a 1-$\sigma$ systematic dispersion
offset can be thought of as a velocity dispersion anisotropy of $\sim
8$\%.} on {\em all} 9 (10) independent points between 8 and 13 kpc to be
consistent with a round halo (4.1$\sigma$ and 1$\sigma$ times smaller
for a \qrho=0.5 halo).  Similarly, in order to infer a round halo, the
thickness of the gas layer (as presented in Fig. 
\ref{fig:derive.the.flattening}) has to change by +3.4$\sigma_{FWHM}$
(+2.5$\sigma_{FWHM}$) for the varying (fixed) inclination case
(+2.1$\sigma_{FWHM}$ and +1$\sigma_{FWHM}$ for a \qrho=0.5 halo). 
Combining these systematic effects, I find that a round halo requires a
2.2$\sigma$ (1.7$\sigma$) change for the varying (fixed) inclination
case and a 1.5$\sigma$ (0.5$\sigma$) change for a \qrho = 0.5 halo.  We
see that it is highly unlikely that the systematic offsets from the
measured values of $\sigma_{\rm gas}$ and $FWHM_{\rm z}$ required for
consistency with a round halo, or even for a halo of intermediate
flattening, are chance occurrences. 

Taking the uncertainties due to the extragalactic radiation field (\Sec
\ref{sec-The.Influence.of.the.Extragalactic.Radiation.Field}) into
account, the required systematic offsets in dispersion and thickness to
infer a rounder DM halo become smaller.  A moderately flattened halo
with \qrho = 0.5 is consistent with the fixed inclination case for all
but the smallest $\Phi_{i,4}$-values, while a round halo would be
inferred if measured dispersion values are systematically decreased by
\onehalf times the dispersion errors.  However, when using the preferred
varying inclination case, a systematic change by 0.9$\sigma$
(1.5$\sigma$) of the measured dispersions and gas layer widths is
required in order to infer a \qrho=0.5 (round) DM halo. 

To summarize, assuming only random errors in inclination, velocity
dispersion, gas layer width and extragalactic radiation field leads to
the conclusion that the DM halo of NGC 4244 is highly flattened: \qrho =
0.2 \pmt 0.1.  An extragalactic radiation field equal to the current
2-$\sigma$ upper limit increases this value to \qrho = 0.3 \pmt 0.2.  A
velocity dispersion anisotropy of 25\% (50\%) would be required to
increase \qrho \ to 0.5 (1.0), essentially independent of the value of
the extragalactic radiation field.

\section{DISCUSSION}
\label{sec-Discussion}

The gas layer width increases exponentially with radius if the potential
is dominated by the stellar disk (e.g., van der Kruit 1988; Paper I). 
For this reason, most authors fit exponential functions to the observed
flaring curves and use simple models for the potential of the galaxy to
derive that the stellar mass-to-light ratio is roughly constant with
values between 50 and 100\% of the maximum disk value (van der Kruit
1981; Rupen 1991; Paper II).  A more detailed model of the potential,
like the one described in Paper I, could be used to re-analyse these
data sets.  However, in order to improve upon the current \MoverL{}
estimates, both the vertical stellar distribution, and the gaseous
velocity dispersion need to be accurately determined.  For all the
extra-galactic systems, the flaring measurements do not extend beyond
the optical disks, so that no information regarding the shape of the
dark halo could be obtained.  With the edge of the stellar disk of the
Milky Way at $\sim$14 kpc (Robin \etal 1992) some of the \HI surveys
extend into the halo of the Galaxy.  The interpretation of this data
however would be difficult because of the uncertainties in the rotation
curve (e.g.  Merrifield 1992) and the gaseous velocity dispersion.

Dissipationless galaxy formation simulations tend to produce triaxial DM
halos (e.g., Frenk \etal 1988; Warren \etal 1992), with an intrinsic
flattening distribution peaked at $\qrho=c/a=0.5$ (\pmt 0.15).  Adding
10\% baryonic matter (gas) and including gasdynamics to these
simulations produces more oblate halos while \qrho \ remains unchanged
(Katz \& Gunn 1991; Udry \& Martinet 1994; Dubinski 1994). 
Observationally, all but the flattest shapes have been reported (for a
summary, see \Sec \ref{sec-Introduction}).  Significantly flattened
halos suggest that the process of galaxy formation is more
dissipational, which might occur if a larger fraction of the initial
density perturbation is baryonic.  Since the flaring can be determined
at lower inclinations as well, possibly at inclinations as low as 60\ad,
the shape of dark matter halos surrounding many galaxies of different
Hubble types can be determined.  It will then be possible to gauge the
significance of the highly flattened dark halo of NGC 4244, and place
more stringent constraints on the process of galaxy formation.

Rich clusters contain $\sim (10 \pm 5) h^{-1.5}$ \% (by mass) hot X-ray
emitting gas (Briel \etal 1992; Mushotzky \etal 1995, and references
therein), with $h$ the normalized Hubble constant ($h$ = H$_0$/(100 \kms
Mpc\rtp{-1})).  It might thus be useful to explore the process of galaxy
formation over a larger range of baryon fractions.  With a total
mass-to-light ratio of $(200 - 300) h$ (Bahcall \etal 1995) for these
rich clusters, the mass-to-light ratio of cluster baryons
(\MoverL{Bar,cl}) equals $(32 - 48) h^{-0.5}$.  Standard Big Bang
nucleosynthesis (BBN) models (e.g., Walker \etal 1991) limits
\MoverL{Bar,BBN} to the range $(11 - 35) h^{-1}$.  From these limits it
follows that for $h \geq 0.5$ essentially all baryons in the universe
are located in rich clusters.  In the region where the cluster and BBN
estimates overlap ($h \geq 0.2$), $\MoverL{Bar} \sim (35 \pm 6) h^{-1}$. 
The dynamical mass-to-light ratios of individual galaxies (Broeils 1992,
1995) range from 4 to 100 $h$, with 80\% of the systems between 6 and 20
$h$.  It is thus possible that the dark halos of individual galaxies
consist mainly of non-baryonic dark matter, but a 100\% baryonic dark
halo is also possible (Gott \etal 1974; Briel \etal 1992; Rubin 1993;
Bahcall 1995; Sackett 1995).  It is not clear however where, and in what
form, these baryons reside in the galactic halos since all plausible
forms of baryonic dark matter seem to be excluded (Hegyi \& Olive 1986).

Pfenniger \etal (1994) reviewed cold, rotationally supported, molecular
hydrogen as a dark matter candidate.  In their model, small high density
molecular ``clumpuscules'' form the building blocks of a highly clumped
interstellar stellar medium (ISM) in which star formation is suppressed
as a result of frequent collisions between the clumpuscules.  Their
model is best developed in the region beyond the optical disk.  Inside
the optical disk the state of the ISM is much more complex as a result
of increased radiation field and metallicity.  With the surface
brightness of the cosmic background radiation equal to $\sim$ 8.5
\LSpcsq, the transition between the two regimes is expected to occur
around $R \sim D_{25}/2$.  The fact that in many galaxies the shape of
the rotation curve due to the gas is similar to the observed rotation
curve (Bosma 1981; Carignan \etal 1990; Carignan \& Puche 1990; Broeils
1992, Chap.  10; Paper II) could then be explained if 3 - 10\% of the
gaseous surface density is in atomic form.  Such might be expected in
the context of clumpuscules hypothesis, or other cold-gas dark matter
models (e.g., Gerhard \& Silk 1995).

Since the self-gravity of the gas layer beyond the optical disk strongly
affects the flaring (cf.  Paper I, Sec.  3.3 and Fig.  8), I investigate
whether the clumpuscules hypothesis is consistent with the flaring data
presented in this paper.  Penny Sackett kindly provided the disk-like
surface density distribution inferred from NGC 4244's rotation curve
using a Keplerian as well as a flat extrapolation beyond the last
measured point ($\Sigma_{\rm tot,Kep}$ and $\Sigma_{\rm tot,Flt}$; see
also Sackett 1995).  The inferred DM surface density is affected by the
way the rotation curve is extrapolated: $\Sigma_{\rm DM,Kep} \approx 86
\exp{(-R/5.2)}$, and $\Sigma_{\rm DM,Flt} \approx 55 \exp{(-R/11.4)}$
\MSpcsq, but is almost independent of the stellar mass-to-light ratio. 
The dark-to-\HI surface density rises in exponential fashion (with scale
length $\sim$ 1.7 \pmt 0.2 kpc) from $\sim$ 10 at the edge of the
optical disk to $\sim$ 100 the edge of the \HI disk.  This increase in
dark-to-\HI ratio is mostly due to the declining \HI surface densities
below column densities of $\sim 2 \times 10^{20}$ cm\rtp{-2}.  Carignan
\& Puche (1990) find a similar effect for NGC 7793 where
$\Sigma_{DM}/\Sigma_{\HI}$ increases exponentially for \HI column
densities below $5 \times 10^{20}$ cm\rtp{-2}.  This contrasts Bosma's
(1981) finding that, for column densities above $\sim 10^{20}$
cm\rtp{-2}, the dark-to-\HI surface density ratio is approximately
constant\footnote{Note that the dark halo surface densities were
calculated in three different ways: Sackett inverts an extrapolated
rotation curve, Bosma uses Nordsieck's (1973) approximation to invert
the rotation curve, while Carignan \& Puche calculate the DM surface
density from a $z$-integration through their best fit isothermal DM-halo
model.}. 

The thickness the clumpuscule disk can be calculated\footnote{The
flaring of the dark disk is calculated self-consistently using the
method described in Paper I (see also \Sec \ref{sec-The.Mass.Model}).}
when a vertical velocity dispersion is assumed.  I find that the
clumpuscule disk has a thickness equal to the \HI layer if the velocity
dispersion of the dark disk is 1.1 (1.3) times larger than
$\sigma_{\HI}$, for $\Sigma_{\rm DM,Kep}$ ($\Sigma_{\rm DM,Flt}$).  With
these dispersions the ``Keplerian dark disk'' is close to being stable
against radial instabilities ($Q = 0.8 - 1.2$ beyond the stellar disk)
while the ``Flat dark disk'' is unstable ($Q = 0.8 - 0.3$) in the 5 - 13
kpc range.  Here I include a correction factor ($1+\frac{2 \pi}{R}
({\rm FWHM}_{\rm gas,z}/2.3) \approx 1.3$) due to the thickness of the disk
(e.g., Pfenniger \etal 1994).  The arguments above suggest that the dark
matter may be in the form of a flaring, stable, self-gravitating dark
disk with a velocity dispersion slightly larger than the measured
dispersion in the \HI.  It would be very interesting to obtain flaring
curves and dispersion measurements for other galaxies to further test
the disk-like dark matter hypothesis.

If an anisotropic velocity dispersion ellipsoid of the \HI rather than a
flattened DM halo is the correct explanation for the data presented in
this paper, then this anisotropy would have implications for the
physical state of the ISM beyond the optical disk.  An anisotropy of the
velocity dispersion ellipsoid can occur if, for example, the collision
time between ``particles'' is large. 

In accordance with van der Kruit's (1981) finding that the dark halo of
NGC 891 cannot be as flattened as the stellar distribution, the axial
ratio of NGC 4244's halo (5:1) is smaller than the axial ratio of
stellar disk (by a factor of $\sim$2 or $\sim$1.4 depending on whether
the vertically distribution is exponential or sech$^2$, respectively). 
It seems that the ``conspiracy of shapes'' (Sackett \etal 1994), i.e. 
the similarity of the shapes of the dark and light distributions, breaks
down for the flattest systems.  Again, more determinations of the shape
of dark halos will make it possible to gauge the frequency of highly
flattened halos like the one of NGC 4244.

\section{CONCLUSIONS}
\label{sec-Conclusions}

As in Paper I, the potential of the global mass distribution, the
self-gravity of the gas included, is used to calculate flaring curves
for a series of models with various stellar mass-to-light ratios and
dark halo flattenings.  Comparing the observed flaring curve with the
various model curves I find that the flaring measurements do not
constrain the stellar \MoverL{} beyond the limits imposed by the
rotation curve.  The shape of the halo on the other hand is well
determined.  The dark matter halo of NGC 4244 is significantly
flattened, \qrho=0.2$_{-0.1}^{+0.3}$ (E9 - E5), albeit less so than the
stellar disk.  The value for \qrho \ reported here is somewhat smaller
than found earlier (Olling \& van Gorkom 1995), mainly as a result of a
better treatment of the gaseous self-gravity and the use of the measured
gaseous velocity dispersion rather than an assumed value.  The flaring
\& velocity dispersion measurements beyond 3 optical scale lengths are
consistent with the dark matter being co-spatial with the \HI disk, a
configuration which is stable against radial instabilities.  Uncertainty
in the determined inclination is the largest source of random errors. 
The assumption that the vertical velocity dispersion equals the measured
tangential dispersion is the most important potential systematic error:
the measured flaring curve is consistent with a round halo if the
anisotropy of the gaseous velocity dispersion ellipsoid increases with
radius.  In that case the vertical dispersion of the gas has to be 50 -
70\% of the measured tangential velocity dispersion at the last measured
point.

\acknowledgments

  I like to thank Jacqueline van Gorkom, John Hibbard and Mike
Merrifield for valuable suggestions to improve this paper.  I also thank
Penny Sackett for providing the disk dark matter curve.  I thank NRAO
for the generous allocation of observing time on which this research is
based.  This research has made use of the NASA/IPAC Extragalactic
Database (NED) which is operated by the Jet Propulsion Laboratory,
CALTECH, under contract with the National Aeronautics and Space
Administration.  This work was supported in part through an NSF grant
(AST-90-23254 to J.  van Gorkom) to Columbia University.

\tfRO

\onecolumn

%
%
\sfRO
\tfRO


\sfRO
\tfRO

\sfRO
\tfRO

\end{document}